\documentclass[twocolumn,10pt]{asme2e}
\usepackage{graphicx}
\usepackage{subcaption}
\usepackage{tabularx}
\usepackage{amsmath}
\usepackage{url}
\usepackage{booktabs}
\usepackage{hyperref}
\usepackage{enumitem}

\confshortname{IDETC/CIE 2026}
\conffullname{the ASME 2026 International Design Engineering Technical Conferences \&\\
              Computers and Information in Engineering Conference}

\confdate{August 23-August 26}
\confyear{2026}
\confcity{Huston}
\confcountry{TX}

\papernum{DETC2026-194316}

\title{A Zero-Shot Multi-Agent Framework for Human-Building Interaction via Programmatic Reasoning}

\author{
Yuqi Wang\textsuperscript{1}, Gulai Shen\textsuperscript{2}, Ali Mehmani\textsuperscript{1}\thanks{Corresponding Author: amehmani@prescriptivedata.io}
\\[6pt]
\textsuperscript{1}Nantum AI, New York, NY 10021\\
\textsuperscript{2}Harvard University, Cambridge, MA 02138
}

\begin{document}

\maketitle    

\begin{abstract}
Large Language Model (LLM) offers opportunities to enhance Human-Building Interaction (HBI) by enabling more direct interactions through intuitive interfaces to complex building systems. These systems can be characterized by the vast amounts of data across multiple formats, the lack of nonconfidential and generalizable information, and the requirement of domain expertise for interpretation. Applying LLMs to domain-specific tasks like HBI presents additional challenges. Limited training data makes traditional fine-tuning approaches impractical. Meanwhile, the opacity of LLM training data requires careful integration of domain knowledge to ensure reliability. Additionally, different LLMs exhibit varying alignment characteristics, suggesting that achieving both natural interaction and technical accuracy requires a multi-agent approach. These challenges highlight the need for innovative approaches to adapt LLMs for specialized domains while maintaining accuracy and user engagement. In this paper, we develop a hierarchical multi-agent framework that utilizes semantic routing and programmatic reasoning to decouple natural language understanding from building analytics. Instead of standard RAG approaches, our system employs a ``Doorman'' mechanism for task decomposition and specialized coding agents that generate executable Python scripts for precise arithmetic. We validate this framework on a dataset from more than 200 commercial buildings. Results demonstrate the effectiveness in providing accurate and contextual responses for diverse users, including stakeholders, from tenants to building managers, across various building system applications.
\end{abstract}

\section{INTRODUCTION}
The complex interplay between humans and their constructed surroundings profoundly influences daily life, health, and overall well-being. As urban populations grow, individuals increasingly inhabit and operate within built structures, spending over 90\% of their time indoors \cite{leech_nelson_burnett_aaron_raizenne_2002}. The most basic interactions can occur through operable windows, shades, thermostats, and lighting switches. The result of interaction can be observed from the building's energy use, interior environmental conditions, occupants' satisfaction, and more \cite{day2020hbi_review}. 

Furthermore, buildings are becoming more complex, transcending from mere protective shelters to embody intricate designs that modulate behavior, comfort levels, productivity, and social dynamics. This also leads to a higher barrier to understanding and altering building behaviors for occupants, especially those in large buildings centrally managed with Building Management and Automation Systems (BMS and BAS) \cite{day2020hbi_review}. With the recent advancement in Artificial Intelligence (AI), specifically Large Language Models (LLM), new interaction methods can be formed to enhance the relationship between humans and the built environment \cite{alavi2019hbi_intro}.

In the remainder of this section, we introduce the concepts of Human-Building Interaction (HBI) and Large Language Models (LLMs) before we combine them to propose the enabled next-level HBI through LLM.


\subsection{Human-building Interaction}

HBI encompasses a wide range of interdisciplinary researches aim to understand how humans interact, adapt, and affect the built environment as well as how the built environment impacts human outcomes and experiences \cite{becerik2022hbi}. Viewing HBI from the perspective of intersecting Human Computer Interaction (HCI), HBI holds its core principle to enable the two-way interaction between humans and the built environment, in addition to the technology involved \cite{alavi2019hbi_intro}.

While building consists of a full lifecycle of the design, construction, operation, maintenance, and eventual demolition or reuse \cite{bldg_lifecycle}, the main focus of HBI has been on the operational/use phase during most of its life, when they are most extensively interacting with humans, including the management teams, engineers, and occupants. The objectives, methods, and subsystems that different people interact with can vary significantly, leading to a wide range of research topics. These include concerns about human impact on building energy consumption and indoor environmental conditions \cite{norouziasl2021hbi_energy)}, as well as building's impact on human health, comfort \cite{kim2023hbi_indoorenv}, and overall quality of life \cite{becerik2022hbi_qualityoflife}.

Further, the interaction can happen both passively and actively through direct or indirect interfaces. Traditionally, when buildings are composed of mechanical and electrical systems, humans interact and control buildings directly through manually opening and closing windows and water valves; turning on and off lights and fans; and adjusting air conditioning systems through pneumatic controls. Later, direct digital controls (DDC) emerged in the 1980s, enabling building engineers and occupants to centrally monitor and control buildings through BMS and BAS \cite{boschbmshistory}. More recently, with the emergence of cloud computing, IoT sensors, smart devices, people are able to interact with buildings through smart home applications and building operating systems \cite{nantum_ai, malkawi2023IoTbldg, apple_home}. With the fast development of AI technology, automated building management now continuously monitors and optimizes buildings for energy efficiency and indoor environmental wellness \cite{wang2024indoor}. However, with more technology, it also becomes less and less intuitive, requiring more and more training and expertise to understand what is happening in buildings, interact with buildings, and influence buildings to make more personalized changes \cite{day2020hbi_review}. 


\subsection{Generative AI to Agentic AI}

LLM has rapidly evolved over the past few years, with significant milestones marking its development. The journey began with early models like Word2Vec\cite{church_word2vec_2017} and GloVe\cite{pennington_glove_2014} in the early 2010s, which focused on word embeddings. This was followed by the introduction of transformer-based architectures, particularly with the release of Google's BERT in 2018\cite{devlin_bert_2019}, which set the stage for context-aware language processing. OpenAI's GPT series, starting in 2018, marked another leap forward, culminating in models like GPT-3 and GPT-4, which demonstrated unprecedented capabilities in generating coherent and contextually relevant text. Meanwhile, various companies released their own large language models, including Anthropic's Claude, Google's Gemini, and Meta's LLaMA, available either for token purchase or as open source.

LLMs have been applied across industries such as healthcare, finance, education, and entertainment. Their ability to understand and generate text enables advancements in automated content creation, virtual assistants, decision support systems, etc \cite{zhao2023surveyllm}. Beyond basic text generation, pre-trained LLMs have been explored to be applied across specific downstream tasks, including programming code fill-in-the-middle tasks \cite{pmlr-v235-gong24f} and specific API calling \cite{patil2023gorilla}, synthesizing 3D scenes \cite{pmlr-v235-hu24g}, arithmetic calculation \cite{pmlr-v235-zhang24bk}, etc., while offering improved efficiency, multilingual support, and enhanced human-computer interaction \cite{zhao2023surveyllm}.

Despite all the progress, applying LLMs to specific domains remains challenging as LLMs are typically trained on large-scale, general-purpose open datasets, often lacking the depth and specificity required for specialized domains \cite{shi2024continuallearninglargelanguage}. These problems can be characterized by three main challenges.

\begin{itemize}[label=\textbullet]
\item There is limited domain-specific training data and a high cost for fine-tuning. In many cases, information specific to the problem is unseen by the model during training or differs from the specific use case in the actual system. While fine-tuning has been traditionally applied to address this gap by retraining several layers or the whole model on smaller, domain-specific datasets \cite{gururangan2020dontstoppretrainingadapt}, it can be computationally expensive and may introduce biases when the available datasets are limited. Moreover, each target system, such as a building, often differs significantly from others, making it impractical to rely on a single fine-tuning process to meet the diverse requirements of all potential users.

\item Domain-specific tasks demand a deep understanding of specialized terms, concepts, and tools. Without such knowledge embedded into or fed to the model, it is likely to get plausible but incorrect outputs from the general models. For example, in coding generation tasks, while code interpreters can enable calculated answers to questions requiring mathematical computation \cite{mirzadeh2024limitationsmath}, and Retrieval-Augmented Generation (RAG) can reference authoritative knowledge outside of original training data \cite{lewis2020retrieval}, challenges remain in designing and integrating these components efficiently and robustly to ensure accurate domain-specific responses.

\item Pre-trained LLMs' performance varies based on training data, model complexity, and prompt design, creating the alignment challenge of matching the right model to the right application. While multi-agent systems \cite{li2024surveyllm}, where multiple specialized agents work collaboratively, have shown promise, coordinating multiple agents introduces complexity and potential failures, requiring robust design to optimize interactions and ensure seamless communication.
\end{itemize}

Prior research has explored several strategies to address these challenges in domain-specific LLM applications. Fine-tuning approaches, such as the adaptive pruning and tuning strategy proposed by Zhao et al. \cite{pmlr-v235-zhao24g}, have shown promise in reducing computational costs, but the fundamental limitation of scarce domain-specific data is still a challenge in fine-tuning. In terms of task decomposition, Li et al. \cite{pmlr-v235-li24ar} developed methods for breaking down complex tasks into semantic sub-tasks, though their approach was constrained by relying on a single LLM rather than leveraging the diverse capabilities of different models. While Du et al. \cite{Du_multiagent_debate} and Crispino et al. \cite{Crispino_zeroshot_agent} demonstrated the potential of zero-shot strategies in multi-agent systems for improved reasoning and logical responses, their work remained untested in specialized domain applications. These limitations in existing approaches highlight the need for a comprehensive solution that combines the strengths of Retrieval-Augmented Generation (RAG), multi-agent systems, and zero-shot learning while specifically addressing domain-specific challenges.



\subsection{AI Agents and Agentic AI Systems for Buildings}

With LLMs, everyone now has the potential to interact more directly and proactively than ever with the ever-smarter and more complicated buildings. Many of the existing works have begun by exploring different ways to implement LLM to provide more user-friendly interfaces for complicated problems related to buildings. The applications range from architectural design and conceptualization, building modeling, building construction, to building code compliance, shown in detail in Table \ref{table:bllm_in_bldg}.

\begin{table*}[htbp]
\caption{LLM Implementations in Buildings}
\centering 
\resizebox{0.8\textwidth}{!}{%
\begin{tabular}{cc}
\toprule
\textbf{Area} & \textbf{Summary of works} \\
\midrule
Building construction & Scheduling, planning, resource allocation \cite{amer2021llmconstruction, prieto2023llmconstruction, saka2023gptconstruction}. \\
                      & Hazard recognition and safety training \cite{uddin2023llmconstruction}. \\
                      & Regulatory compliance and documentation \cite{saka2023gptconstruction}. \\
\midrule
Building Information Modeling (BIM) & BIM model generation \cite{du2024text2bim}. \\
                                    & Building code compliance check and reports \cite{zheng2023llmcodecheck, chen2024automated}. \\
                                    & Retrieving building information \cite{zheng2023llmconstruction}. \\
\midrule
Building Energy Modeling (BEM) & BEM software tutorial and training \cite{su2023hotgpt}. \\
                               & Single object generation and model file modification \cite{zhang2024bemllm}. \\
                               & Running simulation with EnergyPlus models and engine \cite{jiang2024eplusllm}. \\
                               & BEM result visualization and knowledge extraction \cite{zhang2024bemllm}. \\
                               & Co-simulation BIM to BEM process automation \cite{forth2024semantic}. \\
\midrule
Building operation and management & Building regulation interpretation \cite{fuchs2024using}. \\
                                  & Building energy performance consulting and optimization \cite{xiao2024exploring}. \\
                                  & Building system human-in-the-loop control \cite{yang2024llmcontrol}. \\
\bottomrule
\end{tabular}
}
\label{table:bllm_in_bldg}
\end{table*}

By targeting different areas, these works that integrated LLM enabled humans with different roles to interact with buildings more efficiently, proactively, and directly. The required level of expertise is also lowered, while the workload for professionals can be significantly reduced, especially on repetitive and non-creative tasks.

These works, emerging in a very short period of time since models like GPT and Claude became available, further highlight the promising potential for HBI to be elevated to more closely connect humans, buildings, and the technologies around them. However, a structured system to organize and teach LLMs about different aspects of buildings is lacking, as building information is scattered across public or private databases, codes, applications, documents, and the Internet. Further, a consolidated interface is needed to capture human instructions and surface processes and results. Lastly, a significant amount of needs from building managers and occupants remained unsolved by the above-mentioned works as they often require access to a wide range of information in addition to the expertise knowledge, as well as tools like code interpreters and a user-friendly interface.






\section{METHOD AND DESIGN}
\begin{figure*}[htbp]
    \centering
    {\includegraphics[width=0.75\textwidth]{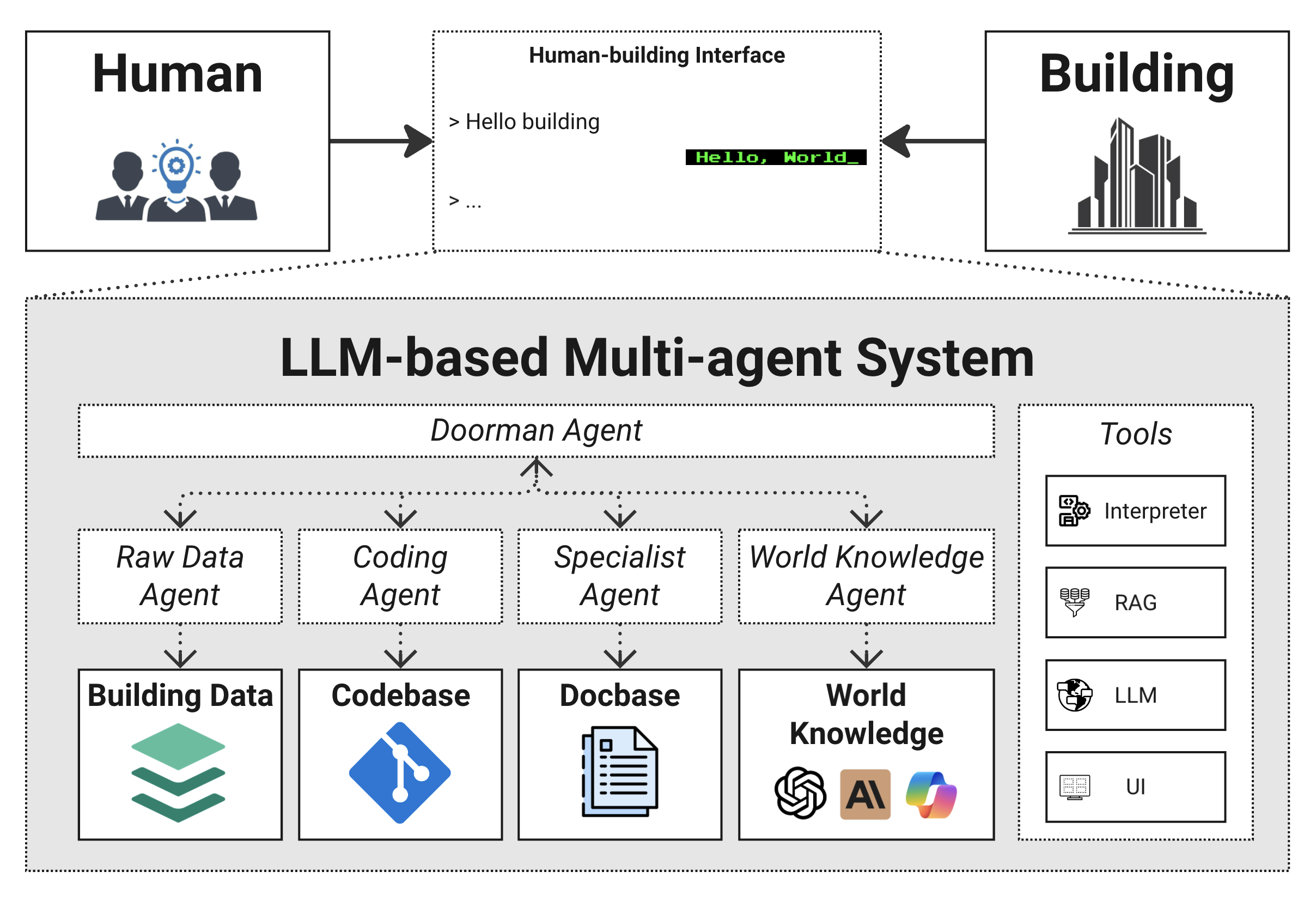}}
    \hspace{1em} 
    \caption{System architecture design for the LLM-based multi-agent system that enables LLM to perform domain-specific tasks using a wide range of data and human building interaction through natural language.}
    \label{fig:sys_design}
\end{figure*}

To address these challenges in the context of Human-Building Interaction, we propose a zero-shot LLM-based multi-agent system framework that integrates real-time databases, code repositories, and technical documents. Our framework specifically tackles: (1) the limited training data challenge through a zero-shot mechanism that eliminates the need for domain-specific training, (2) the domain expertise requirement via RAG integration with building-specific documentation and real-time data, and (3) the alignment challenge through a carefully designed multi-agent collaboration system that matches different types of queries to specialized agents. In the following section, we examine the domain of HBI and its specific requirements for LLM integration, which is the key motivation for the development of our framework. The developed framework, shown in Figure \ref{fig:sys_design}, for the LLM-based multi-agent system comprises three key components: agents, tools, and databases. Agents are categorized as a Doorman agent, a world knowledge agent, coding agents, and specialist agents. The Doorman agent has two primary tasks: (i) splitting the user query into sub-tasks and assigning each to the appropriate agent, and (ii) collecting responses from the agents and providing the final answer. Each agent is equipped with tools such as code interpreters, RAG modules, and LLMs for language-generative tasks. They also access relevant databases, including coding databases, world knowledge databases, and technical documentation, to obtain necessary domain knowledge.

\subsection{Agents}
\paragraph{Doorman agent}
The Doorman agent acts as the central coordinator within the multi-agent system. The agent does not have the ability to directly solve any question, while it holds knowledge about the capabilities and limitations of all other agents. Its primary responsibilities include parsing the user's query and breaking it down into manageable subtasks, and assigning these subtasks to the most suitable agents based on their specialized capabilities. Once the agents complete their tasks, the Doorman agent collects the responses, synthesizes the information, and formulates the final answer to the user's query.

\paragraph{Raw data agent} Raw data agents handle the data retrieval based on the questions and subtasks from the Doorman Agent. It has a thorough understanding of the database structure, types of data stored, metadata information, and data relations. It uses tools like an interpreter and RAG to securely access the database and retrieve required data.

\paragraph{Coding agent} Coding agents specialize in handling programming-related queries. It utilizes tools like code interpreters and domain-specific application programming Interfaces (API) to generate code for answering logical questions.

\paragraph{Specialist agents} Specialist agents are domain experts designed to handle queries that require deep technical knowledge in specific areas. Whether it's engineering design, product, or any specialized field, these agents leverage their access to specialized databases and documentation to provide accurate and detailed responses.

\paragraph{World knowledge agent} The world knowledge agent provides context and general information that draws from a wide-ranging knowledge base. This agent is particularly useful for answering questions related to facts, historical data, and other information that requires a broad understanding or contextual background.

\subsection{Tools}
The system is equipped with a variety of tools to enhance the capabilities of the agents.

\paragraph{Interpreter} Code interpreter allows coding agents to execute and test code snippets in real-time and in the local environment. In our developed system, the code interpreter is a crucial tool to run code that retrieves data, interacts with various APIs, and generates results and plots. It ensures calculations are performed in a consistent and accurate way while allowing flexibility for presenting the results depending on the needs.

\paragraph{Retrieval-Augmented Generation (RAG)} RAG modules enable agents to retrieve information from external databases or documents to generate more accurate and contextually relevant responses. In the developed system, agents prioritize the available databases to search for relevant data, documents, and codes that can significantly improve answers to be more grounded and accurate.

\paragraph{Language Model Models (LLMs)} LLMs are utilized by the agents for natural language processing tasks, such as understanding queries, formulating responses in a human-like manner, etc.

\paragraph{User Interface (UI)} UI enables a chat-based interface for the users to put in questions and get system responses in the form of text, images, downloadable files, and potentially more.

\subsection{Databases}
The agents access several key databases to fulfill their tasks:

\paragraph{Building Data} This includes data related to the buildings and their surrounding environments. The data includes the numerical values related to building operations and equipment, building occupancy, weather conditions, and more. Metadata information is stored to allow agents to understand the type of data, the relationships to each other, and other additional information.

\paragraph{Codebase} These databases contain a wealth of programming resources, such as libraries, APIs, and coding examples, which coding agents use to develop and optimize solutions.

\paragraph{Docbase} Specialist agents rely on domain documentation databases that contain in-depth information, such as scientific papers, engineering manuals, and industry standards.

\paragraph{World Knowledge} These databases include encyclopedic knowledge and other information that support the world knowledge agent in providing accurate and contextually appropriate responses.


\section{APPLICATIONS AND RESULTS}
\subsection{Setup and database}

We tested our method given access to Nantum AI \cite{nantum_ai}, where building data, application documentations, and code repositories are hosted. It is assumed that people ask questions with the right level of access to confidential building information. A chat-based interface developed as a web application using Streamlit is used as the human building interface for testing. The conversation management is done using Chainlit. Answers to the questions are provided in the form of appropriate text, figures, and data files if necessary. For the experimental setup, the LLM models used include GPT-4o, GPT-3.5 turbo, Claude 3.5 haiku, and Claude 3.5 sonnet. The models are chosen based on the agents and their tasks.

\subsection{Question design}

A set of questions is designed for testing, ranging from general knowledge about buildings, real-time conditions of the buildings, to the specific impact of each Energy Conservation Measure (ECM) running in the building. The full list of questions is listed in Table \ref{tab:question_design}. We considered three types of occupant roles, namely tenant, building engineer, and management. For each of the roles, they would have different levels of knowledge and varying interests in various aspects of the buildings. For example, tenants might be more interested in the comfort, while the management team might consider the overall energy consumption and sustainability targets on carbon emissions, and so on.

\begin{table*}[htbp]
\centering
\resizebox{0.9\textwidth}{!}{%
\begin{tabular}{ccc}
\toprule
\textbf{Human Roles} & \textbf{Question Types} & \textbf{Questions} \\
\midrule
Tenant / Occupant & Data Retrieval & What is the current temperature in the building? \\
                  &                & How many people are in the building now? \\
\cmidrule(lr){2-3}
                  & General         & What measures are in place to reduce energy consumption? \\
                  &                & What is the significance of reducing carbon emissions from buildings? \\
\midrule
Building Engineer & Energy & What was the total electric consumption for last month? \\
                  &                     & What was the steam total consumption for the last 7 days? \\
\cmidrule(lr){2-3}
                  & Peak Demand         & What was the peak demand for the last billing period? \\
                  &                     & What is today's peak demand so far? \\
\cmidrule(lr){2-3}
                  & Occupancy           & What time do people start to arrive at the building? \\
                  &                     & What is the maximum occupancy on average for my building? \\
\cmidrule(lr){2-3}
                  & ECM                 & What are the ECMs that can help me reduce energy consumption? \\
                  &                     & What is Nantum automated demand management? \\
\midrule
Management        & Sustainability      & What's the total GHG emission yesterday? \\
                  &                     & What is LL97? \\
                  &                     & How can I lower my GHG emissions? \\
\cmidrule(lr){2-3}
                  & Engineering         & What's the compliance rate of Nantum lunch ramp-up commands last week? \\
                  &                     & How much energy did the ECMs save for my building? \\
                  &                     & How can you achieve zero carbon emission? \\
\cmidrule(lr){2-3}
                  & Property & What's the energy saving in the last billing period? \\
                  &                     & How does the current GHG compare to last year? \\
                  &                     & What's the average daily water consumption in 2019? \\
                  &                     & How is building A compared to building B's average daily electricity consumption? \\
\cmidrule(lr){2-3}
                  & General             & How can Nantum help reduce the GHG emissions? \\
                  &                     & What Nantum applications or ECMs help managing and reducing peak demand? \\
\midrule
\end{tabular}
}
\caption{Designed questions}
\label{tab:question_design}
\end{table*}

\subsection{Results}

In order to evaluate the results, we first quantify the response time, accuracy in numerical results, data retrieved, and documentation retrieved based on the question shown in Table \ref{tab:performance_evaluation}. Next, human responses are gathered from 63.7\% of researchers, 9.1\% of product managers, 9.1\% of building operators/managers, and 18.2\% of occupants. 36.4\% of people have more than 5 years of experience working with buildings. The full questions and answers generated, as well as the user evaluation results, can be found in Appendix \ref{app:full}.

As an example shown in Figure \ref{fig:example_answer}, when asked the question "What time do people start to arrive at the building?", the multi-agent system went through the database to find the appropriate occupancy data and invoked the interpreter for code, figure, and CSV file generation. The resulting response contains the figure and data file that are both downloadable. In addition, the text answer is provided at the bottom. The answer shows that the system clearly understood the question and accessed the right database to provide meaningful and accurate answers with illustrative graphs.

\begin{figure}[htbp]
    \centering
    {\includegraphics[width=0.45\textwidth]{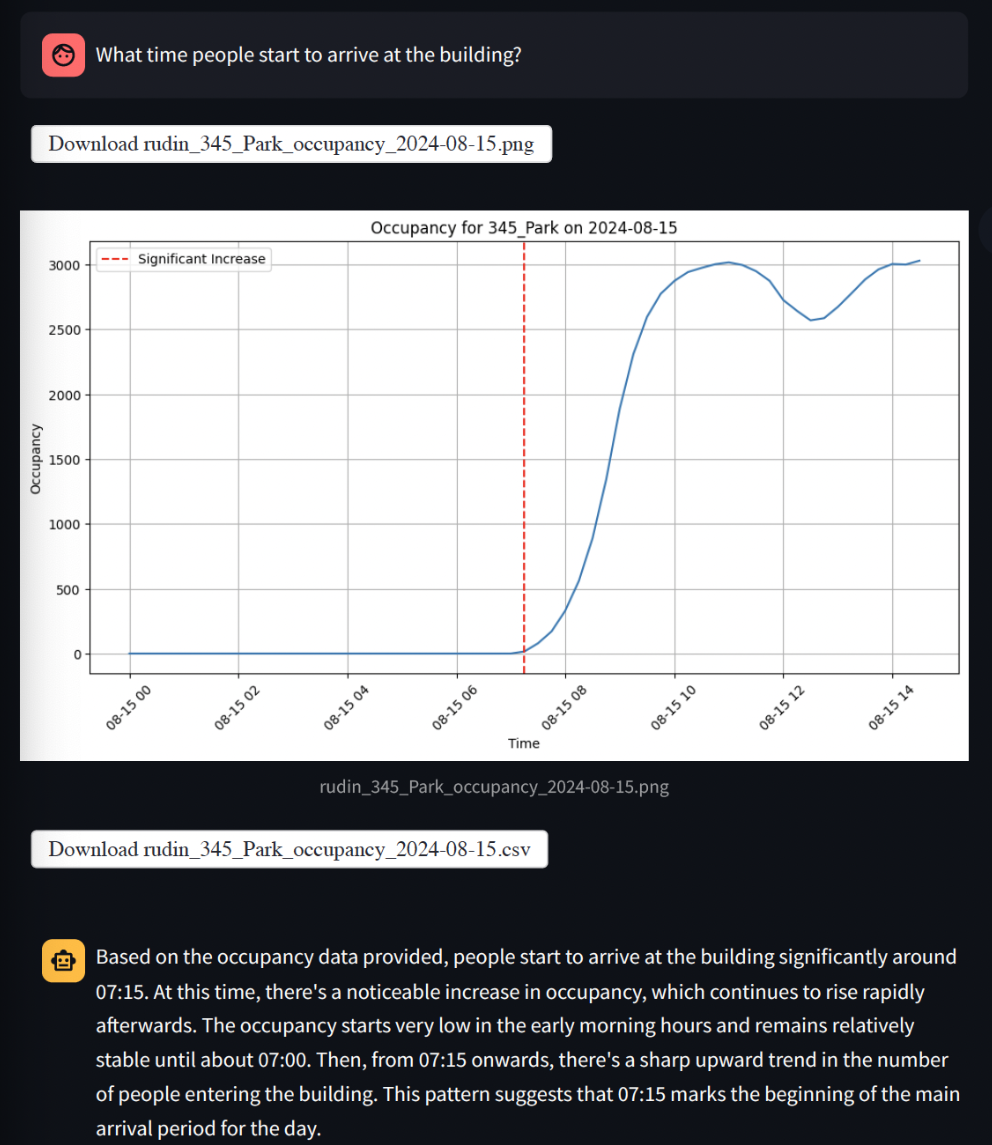}}
    \hspace{1em} 
    \caption{Example question and answer provided by the designed system.}
    \label{fig:example_answer}
\end{figure}

\begin{table*}[ht]
\centering
\resizebox{0.8\textwidth}{!}{%
\begin{tabular}{ccccc}
\toprule
\textbf{Question} & \textbf{Response Time} & \textbf{Numerical Accuracy} & \textbf{Data Access Accuracy} & \textbf{Doc Retrieval Accuracy} \\
\midrule
Q1 & 29s & 100\% & 100\% & 100\% \\
Q2 & 26s & 100\% & 100\% & 100\% \\
Q3 & 15s & N/A & 100\% & 100\% \\
Q4 & 13s & N/A & 100\% & 100\% \\
Q5 & 22s & 100\% & 100\% & 100\% \\
Q6 & 25s & 100\% & 100\% & 100\% \\
Q7 & 24s & 100\% & 100\% & 100\% \\
Q8 & 23s & 100\% & 100\% & 100\% \\
Q9 & 20s & 100\% & 100\% & 100\% \\
Q10 & 33s & 100\% & 100\% & 100\% \\
Q11 & 11s & N/A & 100\% & 100\% \\
Q12 & 15s & N/A & 100\% & 100\% \\
Q13 & 19s & 100\% & 100\% & 100\% \\
Q14 & 30s & 100\% & 50\% & 100\% \\
Q15 & 27s & N/A & 100\% & 100\% \\
Q16 & 53s & 100\% & 100\% & 100\% \\
Q17 & 53s & 100\% & 100\% & 100\% \\
Q18 & 12s & N/A & 100\% & N/A \\
Q19 & 54s & 100\% & 100\% & 100\% \\
Q20 & 26s & 100\% & 100\% & 100\% \\
Q21 & 22s & 100\% & 100\% & 100\% \\
Q22 & 19s & 100\% & 100\% & 100\% \\
Q23 & 15s & N/A & 100\% & 100\% \\
Q24 & 19s & N/A & 100\% & 100\% \\
\bottomrule
\end{tabular}
}
\caption{Quantitative evaluation results for the designed questions}
\label{tab:performance_evaluation}
\end{table*}


\section{DISCUSSION}
The developed system bridges the gap between humans, building, and computer/AI systems by enabling intuitive, language-based communication through the chat-based human building interface. Unlike traditional interaction mechanisms that require specialized knowledge and training, the developed system delivers personalized, actionable answers and insights based on user queries, accessing concrete building information, including data, documentation, and code. The architecture designed is modular and scalable, with flexibility to incorporate new data and switch to different LLM models with ease. At the same time, the multi-agent approach proposed for facilitating human and buildings can also fit other applications. Such an LLM-based system can be a great interpreter to translate human needs to the right ways to retrieve data, write codes, and search documents; on the other hand, it can put data, codes, and technical documents into texts, figures, and more that can be easily understood by different types of users.

It is important to note that the choice of not performing fine-tuning or pretraining comes down to the lack of high-quality labeled data. Instead, we treat the pretrained model as a general-purpose reasoning engine, applying value shaping post hoc rather than altering its internal knowledge. This approach allows efficient use of data and guides the system to specific tasks and capabilities of different agents. The approach also requires thorough documentation for code use when accessing different application APIs and understanding their purposes and limitations. Critically, it also enables the system to adapt to system changes, as we do not need to train the solver each time we move to a new building environment or add/remove new modules.

Furthermore, it is essential to distinguish this architecture from simpler alternatives like a vanilla single LLM or a single-agent ReAct loop. A vanilla LLM baseline would inherently fail on the majority of our tested queries, as it lacks access to the private building telemetry and metadata. Even a single-agent ReAct approach, while capable of tool invocation, often suffers from reasoning drift or alignment failures when navigating the diverse data formats and high technical precision required for HBI. By decoupling high-level task decomposition through the Doorman agent from specialized execution, our framework ensures technical accuracy and prevents the hallucinations that typically plague general-purpose models in specialized engineering domains.

Testing on a range of carefully designed questions, our experiments demonstrate that the system can effectively interpret diverse user queries, access relevant data from structured and unstructured sources, and generate accurate, meaningful responses. For example, queries about real-time building conditions or energy consumption were met with detailed outputs, including text explanations, graphical representations, and downloadable data files. These results validate the system’s potential to make building interactions more accessible, reduce operational complexity, and support decision-making processes across different stakeholder groups. Users generally appreciate the HBI AI chatbot for its responsiveness, accuracy, and ability to provide relevant information efficiently. Many users find it helpful for answering queries quickly and easily navigating complex topics. However, some users express concerns about occasional redundancies or generic responses that lack depth. Others feel that the chatbot sometimes struggles with nuanced or context-dependent inquiries, leading to frustration when detailed or highly specific answers are needed. Additionally, there may be concerns regarding the chatbot's ability to handle more complex multi-turn conversations effectively. Overall, while the HBI AI chatbot is valued for its responsiveness and accuracy, users suggest improvements in contextual understanding, conciseness, and depth of responses to enhance its utility.

Limitations for the current solution exist. First, multi-step sequential and recursive tasks requiring sequential and/or recursive use of different agents cannot be achieved. The doorman agent is currently able to call the appropriate agent for the task, but it cannot arrange orders or assign a specific agent to run multiple times at different stages. Second, the system relies on cleaned and structured datasets. Uncleaned or unstructured data may be called and used, but problems with improper units, missing data, or other issues could cause data loading to fail. Third, a silent error can still cause confident but wrong answers. One missing data point could cause the calculation to potentially be wrong or fail. Furthermore, we have not connected the system to enable actual control of buildings through BMS/BAS. It is a natural next step to enable such a system in combination with interactive reinforcement learning and human-in-the-loop control for building and its subsystems.


\section{CONCLUSION}
In this study, we developed a novel multi-agent system powered by LLMs to enhance HBI. The system developed enables LLM to perform domain-specific tasks that require expert knowledge, interacting with different user types, and utilizing tools. The system enables a more intuitive form of human building interaction and bridges the information gap between occupants, engineers, managers, and the ever-smarter building automation and control systems through direct interactions and an intuitive interface. With a set of carefully designed questions covering different stakeholders and topics, we show that the system is capable of accurately understanding the questions, accessing the correct information, and constructing meaningful and informative outputs. Human building interaction powered by LLM can have a significant impact on the future of smart buildings to be more sustainable, that is, efficient, healthy, and human-centric. 

    

\begin{acknowledgment}

The authors would like to thank Gurpreet Singh, Amir Behjat, Gerald Toge, Gene Boniberger, John Gilbert, John J. Gilbert IV, Evan Torkos, Amit Paul, Lauren Long, Bryce Nielsen, and Gary Chance from Nantum AI, along with Rudin Management Company, for their support and contributions to this project.

\end{acknowledgment}

%
% ...


\bibliographystyle{asmems4}
\bibliography{reference}

\appendix
\section{Survey Instrument and Response Data}
\label{app:full}

The complete survey instrument (including all questions and answer options) and the anonymized user response dataset are available at the following links:

\begin{itemize}
\item Survey instrument: \href{https://forms.gle/Cs764wJz4ahaGi7bA}{Google Form}
\item Response dataset: \href{https://docs.google.com/spreadsheets/d/1E3cIyiDjocHCoWTjXDEbGs95veZcA79ISNvMJROSrE8}{Google Sheet}
\end{itemize}




\end{document}